\documentstyle[aps,prb]{revtex}
\title{NONLINEAR QUANTUM DYNAMICS OF STRONG VIBRATION: RELAXATION JUMPS AND PHONON
BURSTS}
\author{V. HIZHNYAKOV, D. NEVEDROV} 
\address{Institute of Theoretical Physics, University of Tartu,
T\"{a}he 4, Tartu EE2400, ESTONIA. \\
Institute of Physics, Riia 142, Tartu EE2400, ESTONIA.\\
{\rm e-mail: hizh@park.tartu.ee}}
\begin{document}
\thispagestyle{empty}
\maketitle
\begin{small}
\begin{center}
\end{center}
\end{small}

\begin{abstract}
We examine quantum decay of localized vibrations in anharmonic crystal
lattice. The theory which describes two-phonon anharmonic relaxation
can be applied both to local modes associated with substitutional
impurity and to intrinsic local modes (ILM) in perfect
lattices. It is found that for sufficiently high initial excitations
relaxation
of vibrations is non-exponential, it demonstrates explosion-like behavior
at specific stages of evolution.
The course of the relaxation is determined by the initial value of energy, temperature, 
direction of vibrations.
As an example we present the results of calculations of the relaxation of
an odd local (impurity) mode
in a simple cubic lattice and discuss the influence of quantum fluctuations on the 
stability of the ILM in one-dimensional monatomic chain.
\end{abstract}

\section{Introduction}
The localization and energy transport in nonlinear systems have received
much attention recently.
Numerical techniques play an essential role in the investigation of anharmonic
discrete lattices, allowing the variation of model parameters and the observation of
different properties. One of the effects -- the existence of
stationary localized vibrations in a perfect lattice (s.c. intrinsic
local modes (ILM))
\cite{ovchi,dolgov,sivtak,zavt,page} is of particular
interest as it links local lattice dynamics with physics of solitons and
underlines in general the importance of strongly excited local modes.

So far the research of strong anharmonic effects
in lattice dynamics as well as molecular dynamics
simulations of the kinetics of the decay of vibrational
excitations~\cite{burlakov} has been carried out within the frame of classical mechanics.
However, the account of quantum effects may be important, as it leads to new
channels in the decay. In this context
the problem of strong local vibration is also of significant interest,
it gives an example when anharmonic
and quantum effects can be described analitycally with the application
of nonperturbative theory.
One of the results of the corresponding theory
\cite{procest,hizhrev,hizhnev} is an
explosion-like release of energy by the local vibration  with the emission of a burst of
phonons. This release
takes place if the energy of the local mode reaches some definite
critical values. Such a behaviour is new for vibrational systems
and it is a consequence of a mutual interplay between anharmonic and
quantum effects.

The method used in \cite{procest,hizhnev} is based on the assumption that a local mode is
strongly excited initially and it can be considered classically. Phonons are supposed to be
not excited (or
to be in thermal equilibrum).
According to
classical mechanics, the state with
non-excited phonons is stable, since for the phonon coordinates $q_{i}$
at rest ($q_{i} = 0$) the anharmonic
interaction is turned off.
Quantum fluctuations, coming from zero-point vibrations, turn on the
interaction and result the relaxation of the mode. However, the
standard quantum theory,
based on the description of anharmonic interaction as a
small perturbation \cite{ovchi}, cannot be used for the obtaining of
the solution to the problem of relaxation, because the interaction is not
weak (due to the strong excitation of the local mode).

\section{General theory}
Let us consider a local vibration anharmonically interacting with phonons.
The peculiarity of the problem is that
the strong local vibration causes periodic time-dependence of the local force constants
and therefore of the zero-point energy of the
phonon system. This time-dependence results the generation of phonons.
The mechanism of this process has an analogy with
the black hole emission
mechanism proposed by Hawking \cite{hawking} and with Unruh radiation
\cite{unruh}. In all these cases the time dependence of the zero-point
energy causes the transformation of the initial creation and destruction
operators in time. Namely, according to Hawking \cite{hawking}
\begin{equation}
\hat{b}_{i}=\sum_{i}{(\bar{\alpha}_{ij} \hat{a}_{j} - \bar{\beta}_{ij} \hat{a}_{j}^{+})}\,,
\label{eq:hawk}
\end{equation}
where $\hat{a}$ and $\hat{b}$ are initial (incoming) and final
(outgoing) time-dependent destruction
operators.
It means that when there are no incoming particles the number operator of the
$i$-th outgoing state is
\begin{equation}
N_i = <0|\,\hat{b}_{i}^{+}\, \hat{b}_{i}\,|0> = \sum_{j}\, |\,\beta_{ij}\,|^{2}\,,
\end{equation}
i.e. the number of particles created and emitted in a gravitational
collapse can
be determined calculating the coefficients $\beta_{ij}$.
There exist several situations in quantum field theory where the phenomena
similar to black hole evaporation appear. The most famous are:
\begin{itemize}
\item{Pair production in a static electric field revealed by Heisenberg
\cite{heis}. Vacuum of quantum field theory is unstable against
the creation of charged pairs.}
\item{Accelerated systems which become spontaneously excited in 
Min\-kow\-ski-space and accelerated mirrors.}
\end{itemize}

It is also known that the rate of emission of the black hole is very high near
the end of its life. As a black hole emits radiation it loses mass. This in turn
increase the rate of emission. Near the end of its life about
$10^{30}$~erg would be released in $0.1$~s. So the emitting rate of the black hole
grows in time and at some critical moment a large amount of energy will be released
in very short time (explosion). 
We want to point out the analogy of this effect in the system of
phonons in nonlinear crystal lattice.

Cubic anharmonic interaction of the
local mode with crystalline is the sum of terms
$\sim \hat{x}_{i}\,\hat{x}_{i'}\,Q(t)$,
where $\hat{x}_{i}$ and $\hat{x}_{i'}$ are the operators of normal coordinates of crystalline
phonons, $Q(t)$ is the time-dependent classical amplitude of the local mode.
This interaction leads to the following time-dependent Hamiltonian of the
phonon system:
\begin{equation}
\hat{H}_{ph}(t) = \frac{1}{2} \sum_{i}(\dot{\hat{x}}_{i}^{2} + \omega_{i}^{2}\,
\hat{x}_{i}^{2})+\frac{1}{2}Q(t)\sum_{ii'}(e_{i} w e_{i'}) \hat{x}_{i}\hat{x}_{i'}\,.
\label{eq:a}
\end{equation}
Here $\omega_{i}$ is the frequency of the mode $i$,
$(e_{i} w e_{i'})\equiv\sum_{n} e_{in}e_{i'n}w_{nn}$,
$n = 1, 2, \ldots ,n_{0}$, $\sum_{i}e_{in}e_{in'}=\delta_{nn'}$,
$n_{0}$ is the number of con\-figura\-tional coordinates con\-tri\-buting to
an\-har\-monic interaction.

The Hamiltonian (\ref{eq:a}) can be diagonalized as follows \cite{procest}:
\begin{equation}
\hat{H}_{ph}(t) = \sum_{j} \hbar \Omega_{j}(t)\,{\Big
(}\hat{b}_{j}^{+}(t)\hat{b}_{j}(t)\,+\,\frac{1}{2}{\Big )}\,,
\label{eq:f}
\end{equation}
where $\Omega_{j}(t)$ are time-dependent phonon frequencies, and
\begin{equation}
\hat{b}_{j}(t) = \sum_{i}{\big (}\mu_{ij}(t) \hat{a}_{i} + 
\nu_{ij}(t) \hat{a}_{i}^{+}{\big )}\,,
\label{eq:g}
\end{equation}
are time-dependent destruction operators; expressions for $\Omega_{j}$,
$\mu_{ij}$ and $\nu_{ij}$ are given in \cite{hizh}.

The anharmonic interaction considered causes not only time dependence of
phonon
frequencies but also changes of phonon operators in time. The relation (\ref{eq:g})
is analogous to the relation of Hawking (\ref{eq:hawk}) between field operators in different times
of gravitationally collapsing star. This is why the mechanism of a local
mode relaxation is analogous to that of black hole emission: phonons (photons)
are generated because the initial zero-point state $|0>$ is not zeroth
state for the time-dependent destruction operators $\hat{b}_{j}$; there are
phonons with frequencies $\Omega_{j}$ in $|0>$ at the time moment $t\,$;
the number of phonons equals $\sum_{j} N_{j}(t)$, where
$N_{j}(t)=<0|\hat{b}_{j}^{+}(t)\hat{b}_{j}(t)|0>$.

Energy, which is generated in phonon system at the time moment $t$, equals \cite{procest,hizhrev}:
\begin{equation}
E_{ph}(t) =  \sum_{j}
\hbar \Omega_{j}(t) {\Big (}N_{j}(t) + \frac{1}{2}{\Big )} - \sum_{i} \frac{\hbar\omega_{i}}{2}
 =  \frac{\hbar}{4} \sum_{ii'}\,\omega_{i}^{-1} e_{i}^{2} e_{i'}^{2}
{\bigg |}\int_{0}^{t}\,D_{i}(\tau)\,e^{-i\,\omega_{i'}\,\tau}\,d\tau {\bigg |}^2\,,
\label{eq:h}
\end{equation}
(fast oscillating terms are neglected). This energy comes from
the local mode: $E_{ph}(t) = E_{l}(0) - E_{l}(t)$;
$E_{l}(t)\simeq \omega_{l}^{2}
Q_{0}^{2}(t)/2$, $Q_{0}(t)$ is the mode amplitude,
$\omega_{l}$ is its frequency. This relation between the energies $E_{ph}$ and $E_{l}$
gives an equation for $E_{l}(t)$.
Differentiating (\ref{eq:h}) and taking into account that
$dE_{ph} = -dE_{l}$,
one obtains $\dot{E}_{l}(t) \approx - \gamma(t)E_{l}(t)$, where
\begin{equation}
\gamma(t) \approx  \frac{\pi\hbar w^{2}}{4\omega_{l}} \sum_{k=1}^{m+1}
\int_{\bar{\omega}_{k-1}}^{\bar{\omega}_{k}}
\frac{d\omega \rho(\omega) \rho(\omega_{l}-\omega)}{|1 - w^{2} E_{l}(t) \tilde{G}(\omega-\omega_{l})/
{(2\omega_{l}^{2})|^{2}}}
{\Big [}1 - \Theta(E_{k} - E_{l}(t)) e^{-2\Gamma_{k} t}{\Big ]}
\label{eq:gamma}
\end{equation}
stands for the relaxation rate at the time moment $t$
($\bar{\omega}_{0} = 0$, $\bar{\omega}_{m+1} = \omega_{l}/2$), $\rho(\omega) = -\Im{(G(\omega))}$
is the phonon density of states (for $\omega > 0$), $\tilde{G}(\omega - \omega_{l}) =
G(\omega)G(\omega - \omega_{l})$ is the two-phonon Green's function,
expression for $\Gamma_{k}$ is given in \cite {hizhrev,hizhnev}.
The second term in square
brackets in (\ref{eq:gamma}) is important if the initial energy
is close to critical point. When deriving
expression (\ref{eq:gamma}) it was taken into account that for $E_{l} \approx E_{k}$
\begin{equation}
\gamma_{k} \sim \frac{1 - \Theta(E_{k} - E_{l}(t)) e^{-2 \Gamma_{k}t}}{E_{k} - E_{l}(t)}\,.
\label{eq:m}
\end{equation}

\section{Odd local mode in a cubic lattice}
To illustrate the theory we calculate the relaxation of the local mode associated with
a light substitutional impurity atom in a simple cubic lattice.
Within the
approximation of the nearest neighbours interaction the potential operator
has the form
\begin{equation}
\hat{V} = \sum_{\alpha}\sum_{\vec{n}}\sum_{m=1}^{\infty} \frac{1}{m} K^{(m)}_{
\vec{n}_{\alpha}} {\big (} \hat{R}_{\vec{n}_{\alpha}} {\big )}^{m}\,,
\end{equation}
where $\alpha = x,\:y,\:z$ are the directions of crystal axes, $\vec{n} = (n_{x},n_{y},n_{z})$ is
the vector of the lattice sites, $\vec{n}_{\alpha}$ is the vector of the site
nearest to $\vec{n}$ in $\alpha$ direction,
\begin{equation}
\hat{R}_{\vec{n}_{\alpha}} = \sqrt{(R_{0 n_{\alpha}} + \hat{r}_{\alpha
\vec{n}_{\alpha}})^{2} +
\hat{r}^{2}_{\vec{n}_{\alpha}} - \hat{r}_{\alpha \vec{n}_{\alpha}}^{2}}
\end{equation}
is the operator of distance between the nearest neighbours in the
$\alpha$-direction, $R_{0\vec{n}_{\alpha}}$ is the distance between the nearest
neigbour sites,
$\hat{r}_{\beta \vec{n}_\alpha} = q_{\beta\vec{n}} - q_{\beta \vec{n}_{\alpha}}$,
$q_{\beta}$ is the $\beta$-component of the displacement vector $\vec{q}_{\vec{n}}$
of the atom $\vec{n}$,
$\hat{r}_{\vec{n}_{\alpha}}^{2} = \hat{r}^{2}_{x\vec{n}_{\alpha}} +\hat{r}^{2}_{y\vec{n}_{\alpha}}
 + \hat{r}^{2}_{z \vec{n}_{\alpha}}$.
By expanding $\hat{V}$ in the power series of displacement operators one gets
\begin{eqnarray}
\hat{V}  =  \frac{1}{2}\sum_{\alpha,\vec{n}} {\Big(}V_{2\vec{n}_{\alpha}}
\hat{r}^{2}_{\vec{n}_{\alpha}} +
V'_{2\vec{n}_{\alpha} } \hat{r}^{2}_{\alpha \vec{n}_{\alpha}}
 +  V_{3\vec{n}_{\alpha}} \hat{r}^{2}_{\vec{n}_{\alpha}}
\hat{r}_{{\alpha}\vec{n}_{\alpha}} + V'_{3\vec{n}_{\alpha} }
\hat{r}^{3}_{{\alpha}\vec{n}_{\alpha}}{\Big )} +
\ldots\,,
\label{eq:expr}
\end{eqnarray}
where
\begin{eqnarray}
V_{2\vec{n}_{\alpha}} & = & \sum_{m}R_{0\vec{n}_{\alpha}}^{m-2} K^{(m)}_{\vec{n}_{\alpha}}
\,, \nonumber \\
V_{2\vec{n}_{\alpha}}' & = & \sum_{m}R_{0\vec{n}_{\alpha}}^{m-2} K^{(m)}_{\vec{n}_{\alpha}}
(m-2)\,, \nonumber \\
V_{3\vec{n}_{\alpha}} & = & \sum_{m}R_{0\vec{n}_{\alpha}}^{m-3} K^{(m)}_{\vec{n}_{\alpha}}(m-2)
\,, \nonumber \\
V'_{3\vec{n}_{\alpha}} & = & \frac{1}{3} \sum_{m}R_{0\vec{n}_{\alpha}}^{m-3} K^{(m)}_{\vec{n}_{\alpha}}
(m-2)(m-4)\,.
\end{eqnarray}

The potential considered does not take account of the covalent
interaction which leads to the chemical bonding. This (covalent)
interaction can, however, be easily included in calculations by
introducing additional terms of the type $V'_{2\vec{n}_{\alpha}}$
and $V'_{3\vec{n}_{\alpha}}$.

We make calculations for a local mode associated with a vibration of a
light substitutional impurity atom situated at the origin of our reference
frame . In this case
the solutions of
classical equations of motion, corresponding
to
the local mode, satisfy the conditions for
the mode localized on the impurity atom at the site $n=0$. As a consequence,
all mathematical calculations are simplified, since one can neglect
contribitions of other atoms to the classical motion.
This allows us to neglect also the variations of
the frequency of the local mode and of the constants of anharmonic interaction
which are caused by the variation of the amplitude of the local
mode.

Altogether there are 21 operators of Cartesian coordinates (three for each of 7 atoms
includung the central atom and 6 its nearest neighbours) which contribute
to the relaxation of the local mode in this approximation. Therefore,
the interaction Hamiltonian is
\begin{equation}
\hat{H}' = Q_{l} \,\cos{\omega_{l} t}
\sum_{n,n'=1}^{21}V_{3nn'}\hat{q}_{n}\hat{q}_{n'}\,.
\end{equation}
The matrix $\{V_{3nn'}\}$ is determined by the coefficients of the quadratic
operators of cubic anharmonicity in (\ref{eq:expr}) with
\begin{equation}
\hat{r}_{\beta \vec{n}_{\alpha}} = C_{\beta} + \hat{q}_{\beta \vec{n}} -
\hat{q}_{\beta \vec{n}_{\alpha}}\,,
\end{equation}
where $C_{\beta}$ are Cartesian amplitudes of the local mode;
$n$ denotes the numbers of the central atom and its nearest neighbours
and the Cartesian components of their coordinates.
We choose (see Figure~1):
$\hat{q}_{1} = \hat{x}_{0}$,
$\hat{q}_{2} = \hat{y}_{0}$,
$\hat{q}_{3} = \hat{z}_{0}$,
$\hat{q}_{4} = \hat{x}_{1_{x}}$,
$\hat{q}_{5} = \hat{y}_{1_{x}}$,
$\hat{q}_{6} = \hat{z}_{1_{x}}$,
$\hat{q}_{7} = \hat{x}_{-1_{x}}$,
$\hat{q}_{8} = \hat{y}_{-1_{x}}$,
$\hat{q}_{9} = \hat{z}_{-1_{x}}$,
$\hat{q}_{10} = \hat{1}_{1_{y}}$,
$\hat{q}_{11} = \hat{x}_{1_{x}}$,
$\hat{q}_{12} = \hat{z}_{1_{y}}$,
$\hat{q}_{13} = \hat{y}_{-1_{y}}$,
$\hat{q}_{14} = \hat{x}_{-1_{y}}$,
$\hat{q}_{15} = \hat{z}_{-1_{y}}$,
$\hat{q}_{16} = \hat{z}_{1_{z}}$,
$\hat{q}_{17} = \hat{x}_{1_{z}}$,
$\hat{q}_{18} = \hat{y}_{1_{z}}$,
$\hat{q}_{19} = \hat{z}_{-1_{z}}$,
$\hat{q}_{20} = \hat{x}_{-1_{z}}$,
$\hat{q}_{21} = \hat{y}_{-1_{z}}$;
$C = \sqrt{C_{x}^{2}+C_{y}^{2}+C_{z}^{2}}$.

The matrix $w=\{V_{3nn'}\}/V_{30}$ for three directions in the crystal lattice
is given below ($V_{30} \equiv V_{30_{\alpha}}$;
$\kappa = 1 + V'_{30_{\alpha}} /\! V_{30}$).

The operator $H'$ with the given matrices $w$ can be diagonalized
analitycally.
Since the corresponding final analitycal expressions for the
eigenvalues are complicated,
we restrict ourselves in giving their non-zero values for some chosen
parameters $\kappa$ (see Table~1 and Table~2). \\

\[ w^{(1)} = \left ( \begin{array}{ccccccc}
0   &-M_1&M_1 &-M_2&M_2&-M_3&M_3 \\
-M_1&M_1 &0   &0   &0  &0   &0\\
M_1 &0   &-M_1&0   &0  &0   &0\\
-M_{2}^{\top} & 0 & 0 & M_4  & 0 & 0 & 0\\
M_{2}^{\top} & 0 & 0 & 0 & -M_4 & 0 & 0\\
-M_{3}^{\top} & 0 & 0 & 0 & 0 & M_4 & 0\\
M_{3}^{\top} & 0 & 0 & 0 & 0 & 0 & -M_4
\end{array} \right )\,,\;\; \mbox{{\bf [100]},} \]

\[ w^{(2)} = \frac{1}{\sqrt{2}}\left ( \begin{array}{ccccccc}
0&-M_5&M_5&-M_6&M_6&-M_7&M_7\\
-M_1&M_5&0&0&0&0&0\\
M_1&0&-M_5&0&0&0&0\\
-M_{6}^{\top}&0&0&M_5&0&0&0\\
M_{6}^{\top}&0&0&0&-M_5&0&0\\
-M_{7}^{\top}&0&0&0&0&M_8&0\\
M_{7}^{\top}&0&0&0&0&0&-M_8
\end{array} \right ) \,,\;\; \mbox{{\bf [110]},} \]

\[ w^{(3)} = \frac{1}{\sqrt{3}}\left ( \begin{array}{ccccccc}
0&-M_9&M_9&-M_{10}&M_{10}&-M_{11}&M_{11}\\
-M_9&M_9&0&0&0&0&0\\
M_9&0&-M_9&0&0&0&0\\
-M_{10}^{\top}&0&0&M_9&0&0&0\\
M_{10}^{\top}&0&0&0&-M_9&0&0\\
-M_{11}^{\top}&0&0&0&0&M_9&0\\
M_{11}^{\top}&0&0&0&0&0&-M_9
\end{array} \right )\,,\;\; \mbox{{\bf [111]};} \]

\[ M_{1} = \left (\begin{array}{ccc}
\kappa&0&0\\\noalign{\medskip}0&1&0\\\noalign{
\medskip}0&0&1\end{array}
\right ) ,\;\;
M_{2} = \left (\begin{array}{ccc}
1&0&0\\\noalign{\medskip}0&1&0\\\noalign{\medskip}0
&0&0\end{array}\right ), \]

\[ M_{3} = \left (\begin{array}{ccc}
1&0&0\\\noalign{\medskip}0&0&0\\\noalign{\medskip}0
&1&0\end {array}\right ),\;\;
M_{4} = \left (\begin{array}{ccc}
0&1&0\\\noalign{\medskip}1&0&0\\\noalign{\medskip}0
&0&0\end{array}\right ) ,\]

\[ M_{5} = \left (\begin{array}{ccc}
\kappa&1&0\\\noalign{\medskip}1&1&0\\\noalign{
\medskip}0&0&1\end{array}
\right ),\;\;
M_{6} = \left (\begin{array}{ccc}
1&1&0\\\noalign{\medskip}\kappa&1&0\\\noalign{
\medskip}0&0&1\end{array}
\right ), \]

\[ M_{7} = \left (\begin{array}{ccc}
1&0&0\\\noalign{\medskip}1&0&0\\\noalign{\medskip}
0&1&1\end {array}\right ),\;\;
M_{8} = \left (\begin{array}{ccc}
0&1&1\\\noalign{\medskip}1&0&0\\\noalign{\medskip}
1&0&0\end{array}\right ), \]

\[ M_{9} = \left (\begin{array}{ccc}
\kappa&1&1\\\noalign{\medskip}1&1&0\\\noalign{
\medskip}1&0&1\end{array}
\right ),\;\;
M_{10} =  \left (\begin{array}{ccc}
1&1&0\\\noalign{\medskip}\kappa&1&1\\\noalign{\medskip}
1&0&1\end{array}\right ), \]

\[ M_{11} = \left (\begin{array}{ccc}
1&1&0\\\noalign{\medskip}1&0&1\\\noalign{\medskip}
\kappa&1&1\end{array}
\right ). \]

\begin{center}

{\bf Table~1.} Nonzeroth eigenvalues of the matrices $w$, $\kappa = 5$. \\*

\vspace{3mm}

\begin{tabular}{|c|c|c|} \hline
{\bf [100]} & {\bf [110]} & {\bf [111]} \\ \cline{1-3}
$\pm 8.8175$ & $\pm 7.4270$ & $\pm 6.9176$ \\ \hline
$\pm 2.2361$ & $\pm 5.4146$ & $\pm 4.6212$ \\ \hline
$\pm 1.5003$ & $\pm 2.1830$ & $\pm 0.9218$ \\ \hline
             & $\pm 1.2256$ & $\pm 0.5774$ \\ \hline
             & $\pm 0.8570$ & $\pm 0.3825$ \\ \hline
             & $\pm 0.8259$ & $\pm 0.3586$ \\ \hline
             & $\pm 0.7071$ &   \\ \hline
             & $\pm 0.5813$ &   \\ \hline
\end{tabular}

\vspace{5mm}

{\bf Table~2.} Nonzeroth eigenvalues of the matrices $w$, $\kappa = 1.5$. \\*
\vspace{3mm}
\begin{tabular}{|c|c|c|} \hline
{\bf [100]} & {\bf [110]} & {\bf [111]} \\ \cline{1-3}
$\pm 3.1948$ & $\pm 3.8230$ & $\pm 3.9880$ \\ \hline
$\pm 2.2361$ & $\pm 2.1830$ & $\pm 1.7728$ \\ \hline
$\pm 1.2422$ & $\pm 1.6619$ & $\pm 1.0431$ \\ \hline
             & $\pm 1.1114$ & $\pm 0.5774$ \\ \hline
             & $\pm 0.8570$ & $\pm 0.1377$ \\ \hline
             & $\pm 0.7071$ & $\pm 0.1106$ \\ \hline
             & $\pm 0.3364$ &   \\ \hline
             & $\pm 0.1557$ &   \\ \hline
\end{tabular}

\end{center}

\vspace{5mm}

To proceed further we must know the functions of the local densities of states $\rho_{m}(\omega)$. To
find these functions one should specify
the parameters for our lattice and the substitutional impurity atom. For
qualitative estimations, however, it is possible to involve suitable standard expressions for
density functions which in many cases give good approximation.

For qualitative estimations we use the following density functions:
\begin{equation}
\rho_{m}(\omega) = \lambda_{am} \rho_{a}(\omega) +
\lambda_{om}\rho_{o}(\omega)\,,
\label{eq:rho1}
\end{equation}
where
\begin{equation}
\rho_{a}(\omega) =
(3\omega/\omega_{a}^{3})\sqrt{\omega_{a}^{2}-\omega^{2}}
\label{eq:rho2}
\end{equation} is the density function for acoustic phonons and
\begin{equation}
\rho_{o}(\omega)  =  (8/\!\pi(\omega_{o2} - \omega_{o1})^{2})
\sqrt{(\omega_{o2} - \omega)(\omega-\omega_{o1})}\,,
\quad
\omega_{o1} \leq \omega \leq \omega_{o2} \,.
\label{eq:rho3}
\end{equation}
is the density function for optical phonons, $\lambda_{am}$ and
$\lambda_{om}$ stand for the corresponding contribution of the phonon
bands to the $m$ - th density of state, $\lambda_{am} + \lambda_{om} = 1$,
$\omega_{a}$, $\omega_{o1}$ and $\omega_{o2}$ are the limiting
phonon frequencies for acoustic and optical bands respectively.

Formulas (\ref{eq:rho2}) and
(\ref{eq:rho3}) make a correct account of the Van Hove singularities in the
density of states of phonons in the crystal lattice. We should underline,
that the
two-phonon Green's functions $\tilde{G}_{m}(\omega)$ which are relevant to our process,
are less sensitive to details of local dynamics than
the one-phonon Green's function $G_{m}(\omega)$.

In Figures~2-6 the critical behaviour of
the rate of relaxation of a strongly excited local mode is shown.
To simplify calculations we chose a system of units where $\hbar = k =
1$. On the plots time is measured in units $\omega_{a}^{-1}$, energy
is given in units of vibrational quanta $\omega_{a}$, $\kappa$ is a parameter
of interaction.

The time evolution of the local vibration is illustrated
in Figures~2,~4. The corresponding rates of the decay are shown in Figures~3,~5.
As can be seen, the process of relaxation is
determined by
the set of parameters for interaction coefficients and by the initial energy of
the excitation of vibration. Each relaxation jump leads to the release of about 50\% of local
mode energy in several periods of vibration. Depending on the initial
conditions, the most stable
vibrations can be either in [100] or in [111] direction. This effect of
anisotropic relaxation is specific for the mechanism  described.

In Figures~2,~3 the initial energy ($E_{l}(0) = 700$) and the critical values for directions
[110] and [111] are close. Therefore, the vibration in [100] is more stable. The situation is just
the opposite in Figures~4,~5 ($E_{l}(0) = 1800$), the lifetime of the vibrations in [100]
and [110]
is much shorter than that of the vibration in [111] direction.

We carried out a set of calculations for different temperatures. In Figure~6
a comparative behaviour of relaxation for two temperatures is shown.
Generally, for higher temperatures the damping of the local
vibration is faster.

Finally we should pay attention to the following features of the described process of generation
of phonon bursts by strong local vibration:
\begin{enumerate}
\item{A remarkable time delay between the excitation of the local vibration moment and the
phonon burst, generated by this vibration; this delay may be of hundreds or thousands of vibrational
periods, see Figure~6 -- relaxation in direction [111].}
\item{Quasimonochromatic spectrum of phonons, generated in the burst.}
\item{Dependence on the direction of vibration. Separate calculations should be carried out
for each type of the crystall lattice with a correct account of all possible contributions
from vibrations of atoms.}
\end{enumerate}
These effects may be used for the experimental observation of the process.

\section{Quantum instability of intrinsic local modes in anharmonic monatomic chain}
The case of the ILM in a monatomic one-dimensional lattice is of special interest
due to the following features:
\begin{itemize}
\item{The energy $E_{l}$ and the frequency $\omega_{l}$ of the ILM are correlated.
Explosion-like releases of energy will take place if $E_{l}$ exceeds critical energy. Monatomic
one-dimensional chain is the simplest model to investigate whether this condition is fulfilled.}
\item{The one-site Green's function \cite{econ} has only the imaginary part for the range of frequencies $|\omega| <
\omega_{m}$,
where $\omega_{m}$ is the top band frequency. Therefore, the two-phonon Green's function is real; two-phonon
anharmonic damping causes instability of the ILM of the frequency $\omega_{l} < 2\omega_{m}$.}
\end{itemize}

The potential energy operator which includes linear, cubic and quartic terms has the form
\begin{equation}
\hat{V} = \sum_{n} \sum_{r=2}^{4} \frac{K_{r}}{r}
{\big (}\hat{U}_{n+1} - \hat{U}_{n}{\big )}^{r}\,,
\end{equation}
where $\hat{U}_{r}$ is the operator of the longitudinal displacement of the
$n$-th atom from its equilibrium position, $K_r$ are harmonic ($r=2$) and
anharmonic (cubic: $r=3$, quartic: $r=4$) springs. The operators $\hat{U}_{n}$
satisfy the following equations of motion:
\begin{equation}
\frac{\partial^2 \hat{U}_n}{\partial t^2} =
\sum_{r=2}^{4} {\frac{K_r}{M} {\Big [}(\hat{U}_{n+1}-\hat{U}_{n})^{r-1}
- (\hat{U}_{n} - \hat{U}_{n-1})^{r-1}{\Big ]}}\,.
\end{equation}
Here $M$ is the mass of atoms. We suppose that the ILM with frequency
$\omega \le 2 \omega_m$ is excited at the time $t=0$ at the site $n=0$ and its
nearest neighbors ($\omega_m = 2 \sqrt{K_2/M}$).
Anharmonic interactions are
supposed to be weak, satisfying the condition
$a_{0}^{(0)} \ll A_{0}$, where ${a}_{0}^{(0)} \sim \sqrt{\hbar / 2\omega_m M}$ is
the amplitude of zero
vibrations, $A_0 \sim \sqrt{K_2/K_4}$ is
the amplitude of the local mode.

To take account of the intrinsic local mode we introduce the 
operators $\hat{U}_{n}$ in the form
\begin{equation}
\hat{U}_{n} = A_{n}(t) + \xi_{n} + \hat{q}_{n}\,,
\end{equation}
where the "classical" displacements $A_n$ are supposed to be nearly periodic
functions $A_n(t) \simeq a_n(t) \cos{\omega_{l} t}$, $a_{n}(t)$ and
$\xi_{n}(t)$ are slowly time-dependent
amplitudes and shifts satisfying the equations
\begin{eqnarray}
-\omega_{l}^{2} a_{n} & = & \bar{K}_{2}(\bar{a}_{n+1} - \bar{a}_{n}) + 
2\bar{K}_{3}(\bar{a}_{n+1} \bar{\xi}_{n+1} - \bar{a}_{n} \bar{\xi}_{n}) + \nonumber \\
& + &  3\bar{K}_{4} (\bar{a}_{n+1}^{3} / 4 + \bar{a}_{n+1} \xi _{n+1}^{2} -
\bar{a}_{n}^{3} / 4 - \bar{a}_{n} \bar{\xi}_{n}^{2})\,, \nonumber \\
0 & = & \bar{K}_{2} (\bar{\xi}_{n+1} - \bar{\xi}_{n}) + 
\bar{K}_{3}(\bar{a}_{n+1}^{2}/2 +\bar{\xi}_{n+1}^{2} - \bar{a}_{n}^{2}/2
- \bar{\xi}_{n}^{2}) + \nonumber \\
 & + & 3\bar{K}_{4} (\bar{a}_{n+1}^{2} \bar{\xi}_{n+1}/2
+ \bar{\xi}_{n+1}^{3} -\bar{a}_{n}^{2}\bar{\xi}/2 - \bar{\xi}_{n}^{2}) \,, 
\end{eqnarray}
where we neglected the contribution of higher harmonics $\sim \cos{3\omega_{l} t}$,
$\cos{5\omega_{l} t}$, $\ldots$ \cite{dolgov,page,sivtak} 
and introduced for convenience new variables:
$\bar{a}_{n} = a_{n} - a_{n-1}$,
$\bar{\xi}_{n} = \xi_{n} - \xi_{n-1}$.

Slow variations of $a_n$ and $\xi_{n}$, which come from quantum effects, are described
by time dependent operators $\hat{q}_{n}(t)$, which satisfy the
equations
\begin{equation}
\frac{d^2\hat{q}_n}{dt^2} = W_n(t) (\hat{q}_{n+1} - \hat{q}_{n})
- W_{n-1}(t) (\hat{q}_n - \hat{q}_{n-1})\,.
\end{equation}
Here $W_{n}(t) = K_{2} + v_{n}^{(2)} + 2v_{n}^{(3)} \cos{\omega_{l}t}$, where
$v_{n}^{(2)} = 2K_{3} \bar{\xi}_{n} + 3K_{4} (\bar{a}_{n}^{2}/2 + \bar{\xi}_{n}^{2})$
and
$v_{n}^{(3)} = K_{3} \bar{a}_{n} + \frac{3}{2} K_{4} \bar{a}_{n} \bar{\xi}_{n}$
determine the change of springs caused by the ILM. The terms $\sim K_{3}(\hat{q}_{n} -
\hat{q}_{n-1})^{2}$,
$K_{4} A_{n} (\hat{q}_{n} - \hat{q}_{n-1})^{2}$ and $K_{4}(\hat{q}_{n}-\hat{q}_{n-1})^{3}$ are
supposed to be small and they were neglected. Taking account of these terms is important
if $\omega_{l} > 2\omega_{m}$,
when the two-phonon decay under consideration is forbidden by the energy
conservation law.

The Green's function of a perturbed lattice $G_{m}$ is expressed via the Green's function
of a perfect lattice $G_{m}^{(0)}(\omega)$:
\begin{equation}
G_{m}(\omega) = \frac{G_{m}^{(0)}(\omega)}{1-v_{0}^{(2)}G_{m}^{(0)}(\omega)}\,.
\end{equation}

The rate of decay of the ILM is determined by (\ref{eq:gamma}) with $\rho(\omega) = 2\pi^{-1}\Im{(G(\omega))}$.
The properties of the ILM depend essentially on the parameter $K_{3}^{2}/K_{2}K_{4}$.
According to our estimations, for realistic systems it has the value betweem $1.2$ and $2.1$.
Relaxation jumps take place if the denominator in (\ref{eq:gamma}) approaches zero
(if $|v_{m}G(\omega_{l}/2)| = 1$).

Further analitycal and numerical calculations
based on the real parameters
of interaction have shown that the condition
$|v_{m}G(\omega_{l}/2)| = 1$ is fulfilled for
the cases when both the central bonds and the next to central bonds are accounted.
Therefore we may conclude that intrinsic local modes on perfect anharmonic chain are not
stable. For the typical interaction parameters derived from rare gas solids potentials
this conclusion is valid for the frequencies of the ILM
lying on the interval
$1.2 \omega_{m} \ldots 1.7 \omega_{m}$.

\section{Conclusion}
We showed that quantum (and thermal) effects lead to the creation of extra channels in
the decay of a strong local vibration (intrinsic or associated with an impurity) with the
time-dependence
being strongly nonexponetial.
At very large energies the rate of the decay is relatively low. It increases
during the course of the relaxation, at critical points it gets extremely high values.
Then the rate decreases and after that increases again until the energy reaches next critical point.
The energy of the local mode drops in a short time at each critical point and a burst
of quasimonochromatic phonons is generated. The relaxation rate is higher for
higher temperatures, although the values of critical energies do not depend on
temperature. We also showed that the law of relaxation depends strongly on the
directions of vibration and its behaviour is different for different
initial energies. Strong dependence of the relaxation on the directions
of vibration obtained for cubic crystals is remarkable, since in the case
of the existence of chemical bonds the most stable are vibrations in the
direction of the bonds. This effect may have an important value for
chemical reactions and for the mechanism of defects formation in solids

It follows from formulae (\ref{eq:gamma}) and (\ref{eq:m}) that the damping rate $\gamma$ of the local
mode is strongly enhanced (diverges in our approximation as $\;\sim\,|t-t_{k}|^{-1/2}$) if the mode energy
approaches (at $t=t_{k}$) one of the critical energies $E_{k}$. This enhancement of $\gamma$
is associated with the generation of quasimonochromatic phonons. These phonons are
emitted in pairs: one phonon with the frequency $\omega_{k}$ and another with
the frequency $\:(\omega_{l}\,-\,\omega_{k})$. As a result the decay of the strongly excited
local mode is highly non-exponential: it has step-wise jumps near critical energies.

At a final stage of decay the relaxation becomes exponential, i.e. the rate
of decay is constant. In weak coupling limit the theory presented also gives
the constant rate of decay. This is similar to the result given by perturbation
theory \cite{klemens} and to the numerical simulations \cite{burlakov}
of classical anharmonic chain.

For experimental observation of the relaxation jumps both,
quasimonochromatic spectrum of generated phonons and the time delay
between creation of the strong vibration and the jumps can be used.
Strong local vibrations can be excited e.g. by light or by high
energy particles. Excitation by light takes place e.g. in
luminescence centers in ionic crystals with small quantum yield
of emission; here, due to the nonradiative transitions, the
electronic excitation is abruptly transformed to the local
vibrations. High energy particles create in crystals focusons
which, in turn, efficiently pass their energy to impurities
by excitation of their vibrations.

\section*{Acknowledgement}
Supported by the Estonian Science Foundation Grant No.~369

\newpage
\noindent
{\LARGE {\bf Figure captions}}

\bigskip
\noindent
{\bf Fig.~1}
Displacements which contribute to the relaxation of the central atom. The
central (impurity) atom and its six nearest neighbours are shown. 

\bigskip

\noindent
{\bf Fig.~2}
Relaxation of the energy of the local mode in directions [100] - thick
line,
[110] - thin line, [111] - dotted line. $\kappa = 5$, $E_{l}(0)=700$.

\bigskip

\noindent
{\bf Fig.~3}
Rate of the relaxation of the local mode in directions [100] - thin line,
[110] - thick line, [111] - dotted line. $\kappa = 5$, $E_{l}(0)=700$.

\bigskip

\noindent
{\bf Fig.~4}
Relaxation of the energy of the local mode in directions [100] - thick line, [110] - thin line,
[111] - dotted line. $\kappa = 5$, $E_{l}(0)=1800$.

\bigskip

\noindent
{\bf Fig.~5}
Rate of the relaxation of the local mode in directions [100] - thick line, [110] - thin line,
[111] - dotted line. $\kappa = 5$, $E_{l}(0)=1800$.

\bigskip

\noindent
{\bf Fig.~6}
Relaxation of the energy of the local mode in direction [111], \\
$T=0$ - thick line, T=0.85 - thin dotted line. $\kappa = 5$, $E_{l}(0)=700$.

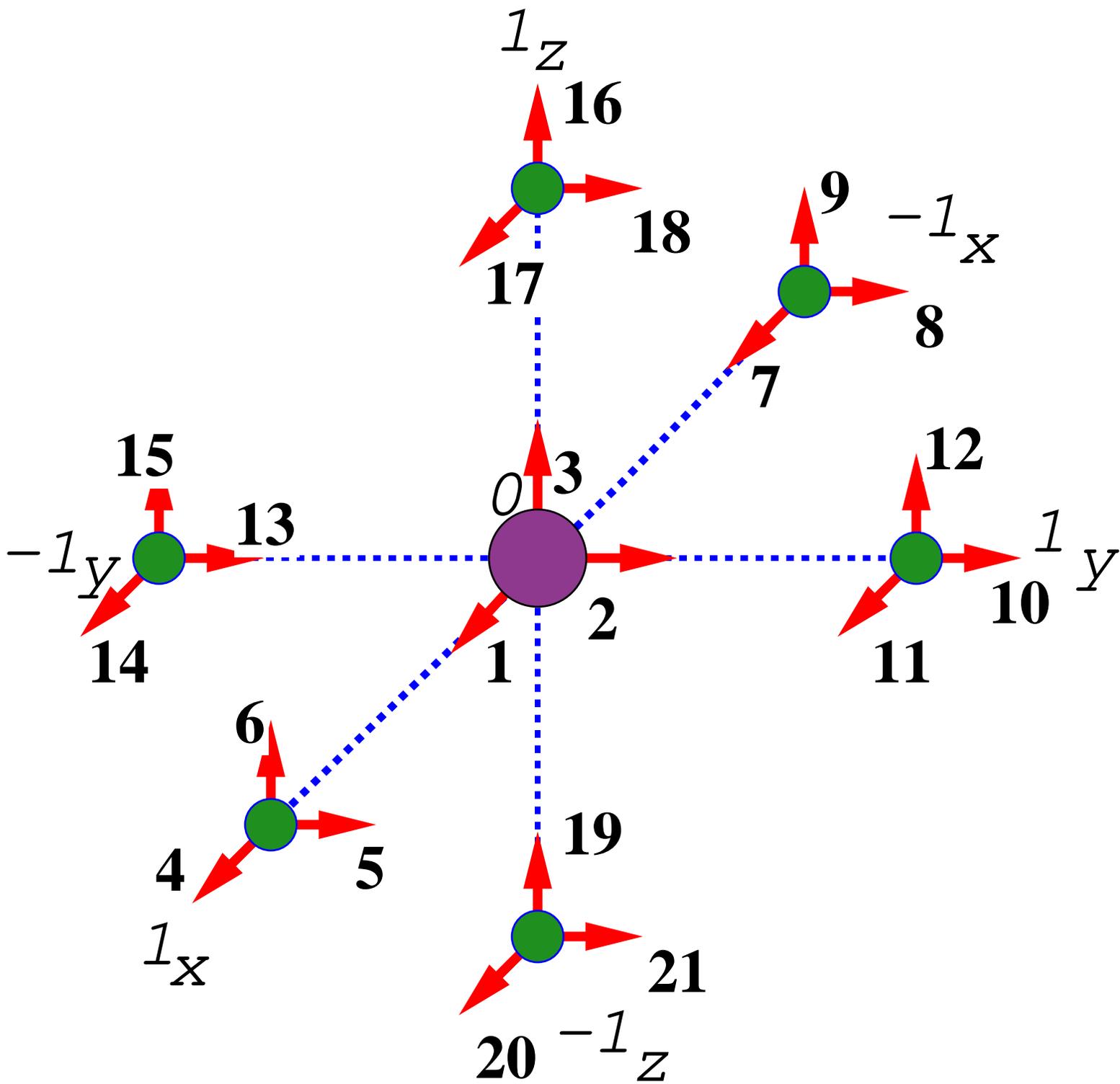
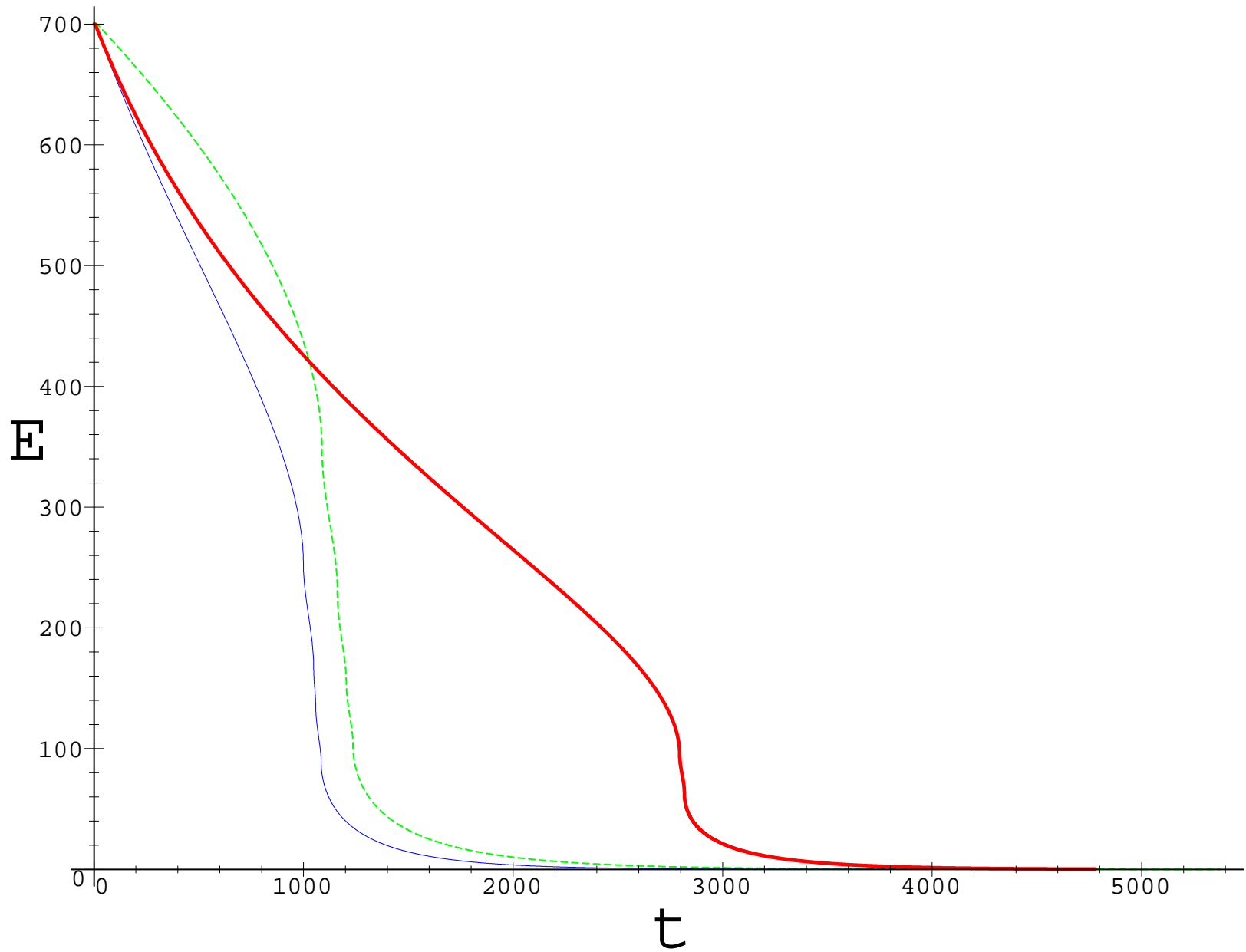
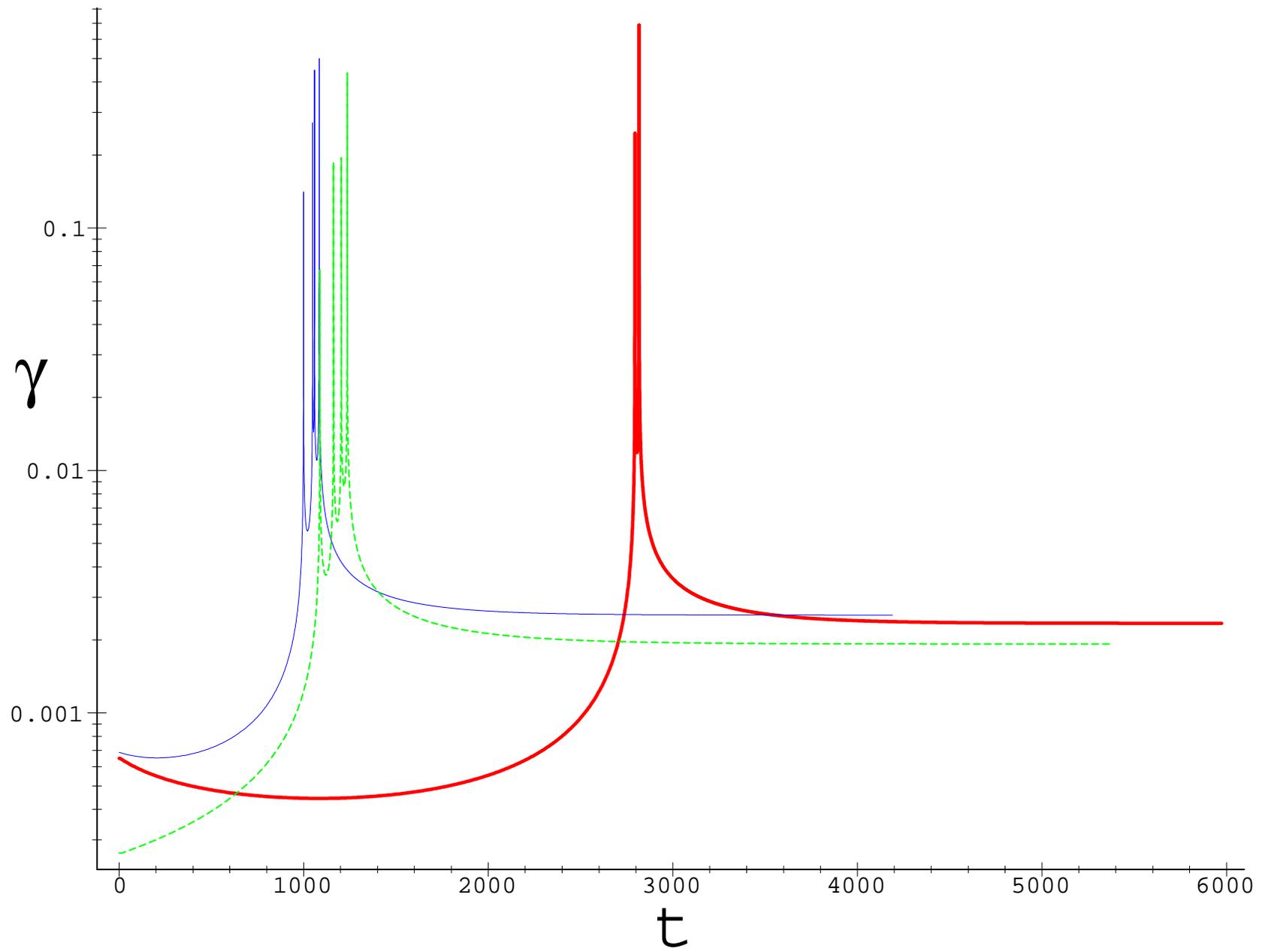
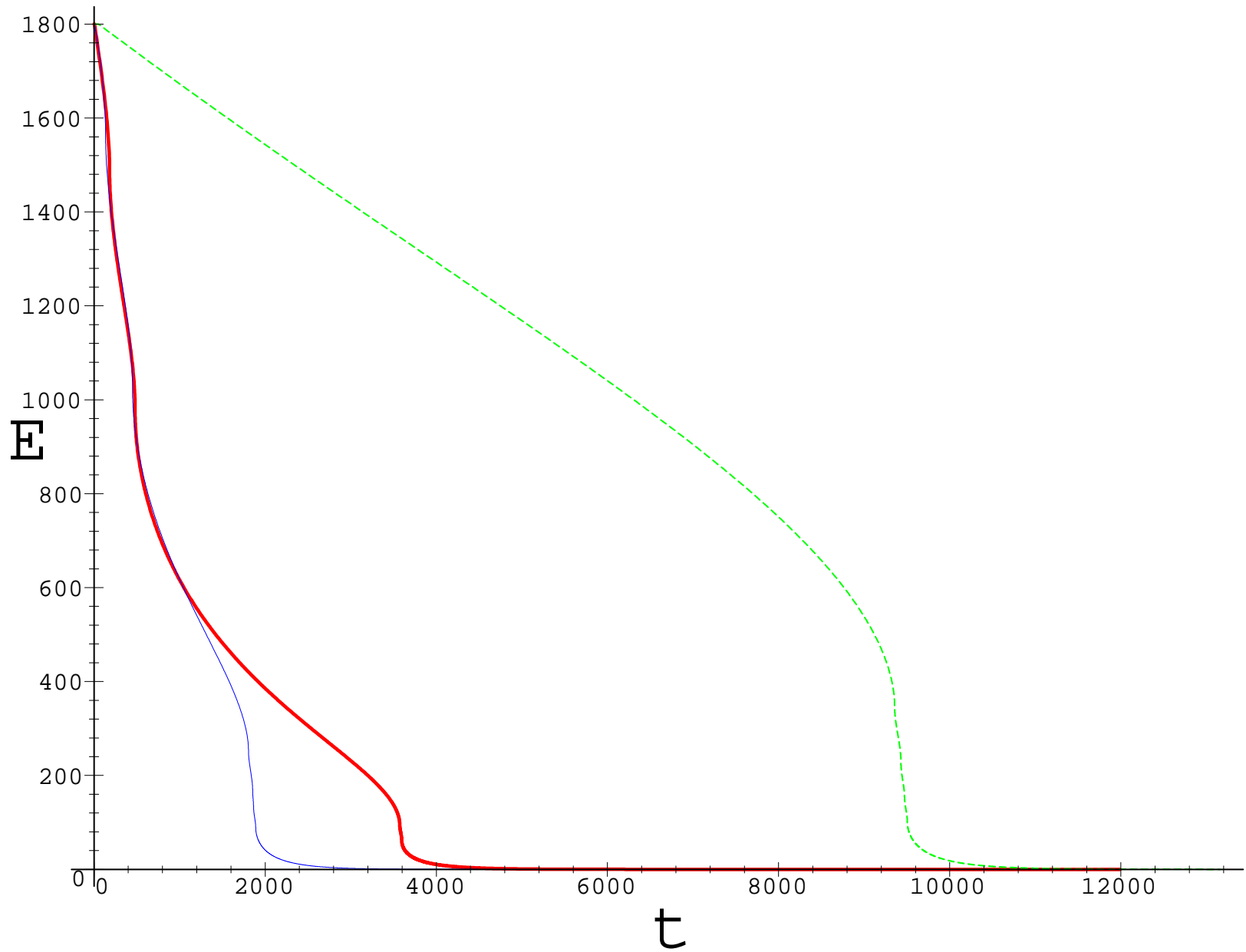
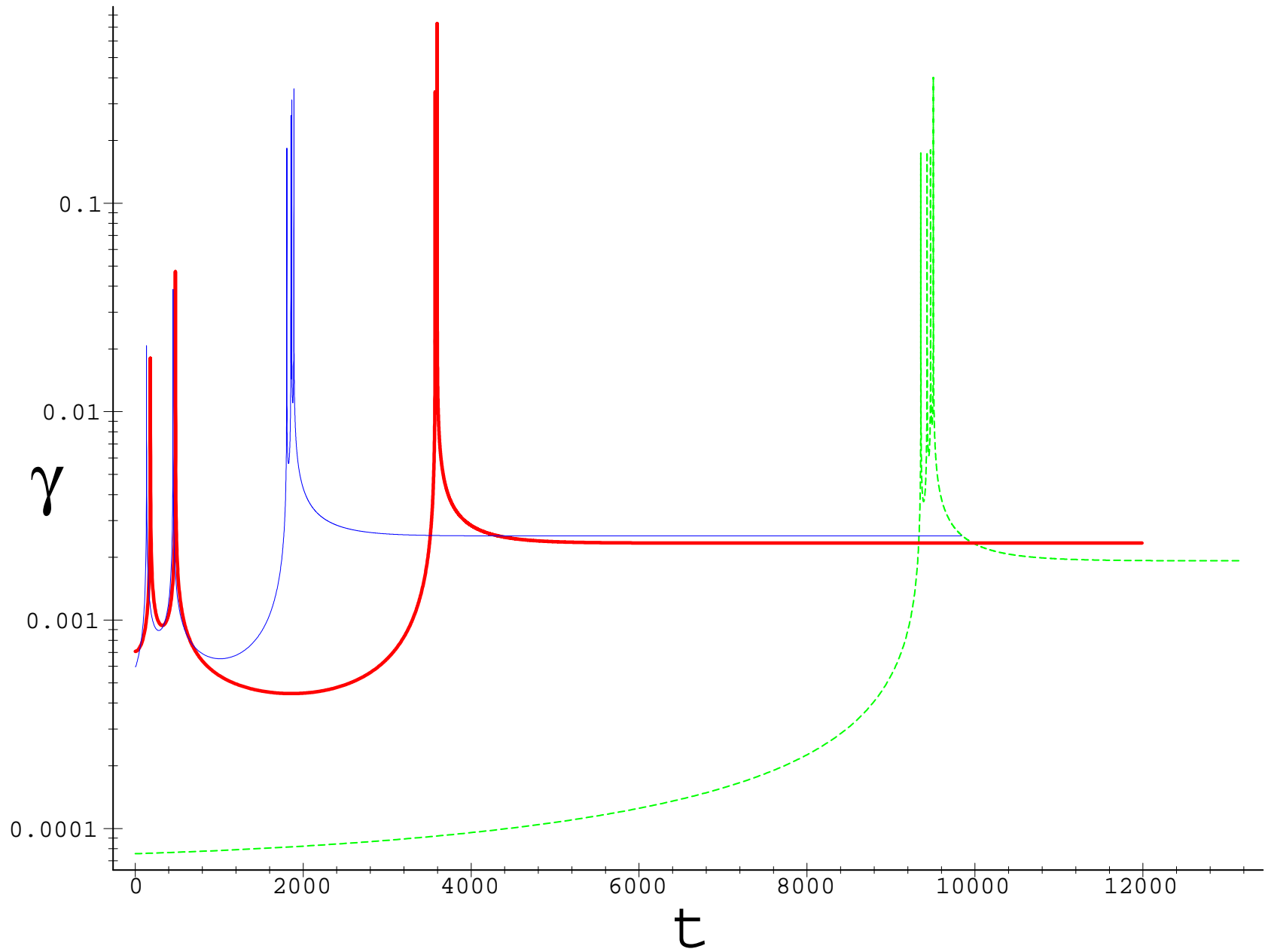
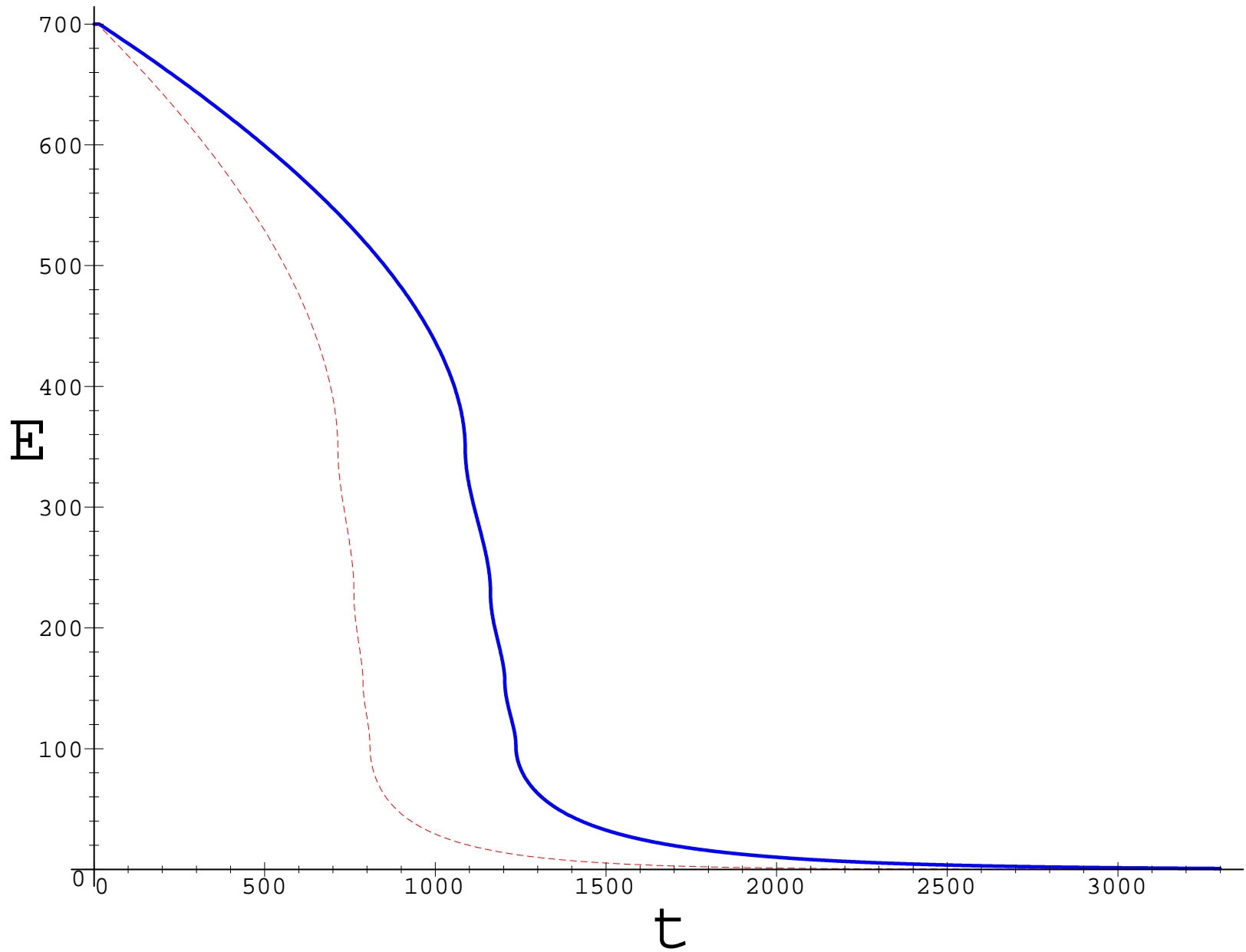
\end{document}